\documentclass[12pt]{article}

\usepackage{amssymb,epsfig}

\graphicspath{{figs/}}

\newcommand{\zeplin}{ZePLiN}
\newcommand{\ul}{\underline}

\begin{document}

\begin{center}
\baselineskip=24pt

{\Large \bf Measurements of muon flux at 1070 metres vertical depth in the Boulby underground laboratory}

\vspace{1cm}

\large
{
M.~Robinson$^{a,}$\footnote{Corresponding author: Matthew Robinson, 
E-mail: matthew.robinson@sheffield.ac.uk}, V.~A.~Kudryavtsev$^a$, R.~L\"uscher$^b$,
J.~E.~McMillan$^a$, P.~K.~Lightfoot$^a$,
N.~J.~C.~Spooner$^a$, N.~J.~T.~Smith$^c$,
I.~Liubarsky$^b$ 
}
\vspace{1cm}

$^a$ 
Department of Physics and Astronomy, University of Sheffield,
Sheffield S3 7RH, UK

$^b$ Blackett Laboratory, Imperial College of Science, Technology and
Medicine, London SW7 2BZ, UK

$^c$ Particle Physics Department, Rutherford Appleton Laboratory,
Chilton, Oxon, OX11 0QX, UK

\vspace{1cm}

\begin{abstract}
Measurements of cosmic-ray muon rates and energy deposition spectra
in a one tonne liquid scintillator detector at 1070 metres vertical depth in the
Boulby underground laboratory are discussed.  In addition,
the simulations used to model the detector are described.  The
results of the simulations are compared to the experimental data and
conclusions given.  The muon flux in the laboratory is found to be 
(4.09$\pm$0.15)$\times$10$^{-8}$~cm$^{-2}$s$^{-1}$.
\end{abstract}

\end{center}

\vspace{0.2cm}
\noindent {\it Key words:} Underground muons,
Dark matter, Muon flux, Neutron background.

\noindent {\it PACS:} 96.40.Tv, 25.30.M, 95.35

\vspace{0.2cm}
\noindent Corresponding author: Matthew Robinson,
Department of Physics and Astronomy, 
University of Sheffield, Hicks Building, Hounsfield Road, 
Sheffield S3 7RH, 
Tel: +44~(0)114 2223553, 
Fax: +44~(0)114 2728079, 
E-mail:~matthew.robinson@sheffield.ac.uk

\pagebreak

\section{Introduction}

Dark matter WIMP (\ul{W}eakly \ul{I}nteracting \ul{M}assive \ul{P}article)
experiments attempt to identify WIMP induced events in ultra low background
detectors.  The dominant backgrounds in such experiments cause electron
recoils within the detectors.  Techniques such as pulse shape
discrimination in a scintillator are used to distinguish between these
events and the nuclear recoil events expected from WIMP interactions.
Neutron interactions also cause nuclear recoils and are not
distinguished from WIMP interactions by these methods.  Alternative
systems must therefore be used to suppress the neutron background in
these experiments. 

Neutrons in underground laboratories are
primarily due to ($\rm\alpha$,n) reactions caused by $\rm\alpha$-decay of U/Th
traces in the rock and detector materials.  Such neutrons typically
have energies of a few MeV.  Passive neutron
shielding of an appropriate thickness around the detectors may be used
to suppress these neutrons, increasing the sensitivity of the experiment.

Neutrons are also produced by muon interactions in the rock and
detector materials.  These neutrons may have much higher energies, the
spectrum extending into the GeV range.  Passive neutron shielding is
not sufficient to suppress such neutrons. 

Muon-induced neutron background may be suppressed using an active muon
veto around the detector so that neutrons accompanied by a muon signal
in the veto may be
rejected.  This will be necessary for experiments to achieve
sensitivities of less than 10$^{-8}$~pb in terms of WIMP-nucleon
interaction cross-section.  The design of such vetos will require a good understanding
of the energy spectrum and flux of muons passing through the
laboratory.  

Since the rock density and composition at Boulby
are not well known, a direct measurement of the muon flux in the
laboratory was performed to 
provide normalisation to simulations of muon-induced neutrons.
A detector normally used as a gamma background veto for the \zeplin~I
(\ul{Z}on\ul{e}d \ul{P}roportional Scintillation in
\ul{Li}quid \ul{N}oble Gases) \cite{bib:zeplin}
dark matter experiment was used to measure this flux.

In this paper, we will describe the measurements made with the
detector.  We will also describe the simulations of the muon spectrum
in the laboratory and of muon interactions in this detector.  Finally, 
we will discuss the analysis of these data to
calculate the muon flux in the experiment and thereby the effective
depth of the laboratory for muon suppression in terms of metres of
water equivalent.

\section{Experimental setup}

\begin{figure}[htb]
\begin{center}
\epsfig{figure=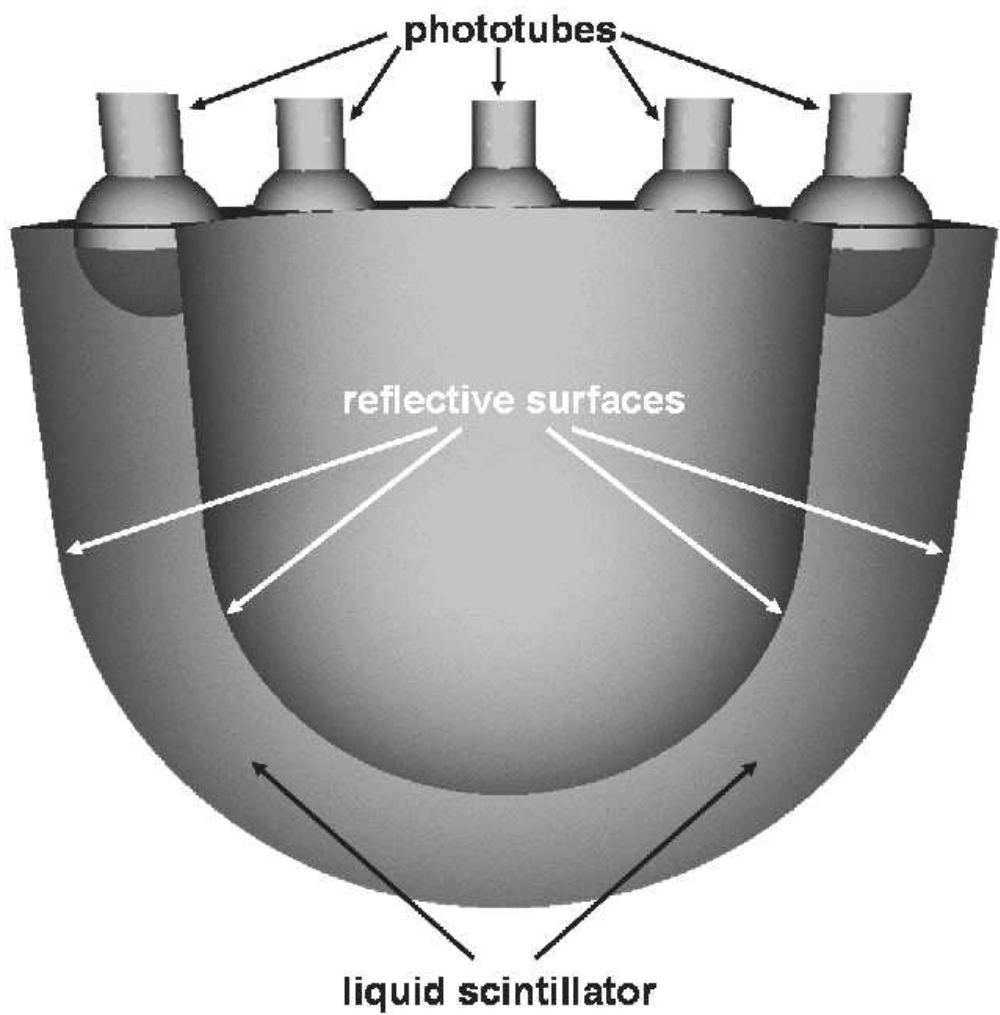,width=15cm}
\caption{Cross sectional diagram of \zeplin\ I gamma veto scintillator
detector which was used to measure muon rate.  The detector has ten
phototubes but only five are shown here}
\label{fig:z1veto}
\end{center}
\end{figure}

The \zeplin\ I veto detector (Figure \ref{fig:z1veto}) is a hollow structure
surrounding the main \zeplin\ I experiment on five sides.  The lower part
of the detector is hemispherical in shape with an inner radius of 0.35~m
and an outer radius of 0.65~m giving a wall thickness of 0.3~m.  The
upper part is cylindrical in shape with the same inner and outer
radii and a height of 0.6~m.  The liquid scintillator within has
a mineral oil base containing approximately 25\% Phenyl-o-xylylethane
(PXE) supplied by Eljen Technology under the product name EJ-399-01 (See
Table \ref{tab:scintillator} for specifications).  This was
chosen primarily for its high flash point
due to safety constraints on materials used at
Boulby.  The volume of the detector
is 1.05~m$^{3}$, giving a total mass of liquid scintillator of 0.93 tonnes.  The
scintillator is viewed from above by ten 20~cm hemispherical
phototubes (Electron tubes \cite{bib:etubes} model 9354KA).  
The inside surfaces of the detector are covered in
aluminium with a coefficient of reflection greater than 0.9 to maximise light collection. 

No specially designed DAq (\ul{D}ata \ul{A}c\ul{q}uisition) system was
used for the muon measurements.  Instead,
the DAq systems developed for the UK dark matter searches were used.
The UK Dark Matter Collaboration (UKDMC) has recently switched DAq from a system employing
digital oscilloscopes read into a G3 Macintosh computer through a GPIB
(\ul{G}eneral \ul{P}urpose \ul{I}nterface \ul{B}us) system
using LabView software, to an Acqiris Compact PCI \cite{bib:acqiris}
digitiser system
read into a PC under custom written Linux software.  Both of these
systems were used to collect muon data in order to examine possible
DAq related systematic errors.  Both DAq systems output
digitised pulses to binary data files.  The pulses stored in the
binary files were analysed using custom written data analysis
software.  The pulses were analysed in terms of amplitude in mV and in
terms of integrated area in mV$\times$ns.  The two analysis methods
produced similar results.  The dynode chains on the
detector PMTs were intended to provide a simple low-level trigger
rather than linear energy measurement and show
pronounced non-linearity effect for high amplitude pulses.  This effect was
quantified by measuring the PMT signal amplitude from fast LED pulses over
a range of amplitudes and the muon data were corrected accordingly. 

The $\rm\gamma$ emission lines at 1275~MeV and
1333~MeV from a $^{60}$Co were used to collect calibration data for the detector.  These
data were analysed using the same techniques as for muon data.  In addition, the
simulations discussed in Section \ref{sec:sim} showed a peak at around
60~MeV which also appeared in the data and was used as a calibration
point.  This peak is due to muons which pass through the bottom part of
the detector and therefore have a track length inside the detector of
30~cm.  
These two calibration points were found to agree in terms of the ratio
of pulse amplitude to deposited energy.

\section{Simulations}
\label{sec:sim}

The simulation code MUSIC \cite{bib:music}, was used to calculate
the angular distribution and energy spectrum of muons expected at various
depths underground.  This was achieved by starting with the muon angular
distribution and energy spectrum at sea level 
and then propagating
these muons through either Boulby rock, or standard rock to various
depths. The muon energy spectrum at various zenith angles at sea level
was taken according to the parameterisation by Gaisser
\cite{bib:gaisser}, modified for large zenith angles \cite{bib:flux2}
with the best fit values for normalisation and spectral index obtained
by the LVD experiment \cite{bib:flux2,bib:flux3}.  Note that the LVD
results also agree with the parameterisation obtained in the MACRO
experiment \cite{bib:flux1}.
Standard rock is
defined to have atomic number 11, atomic mass 22 and density 2.65~g/cm$\rm^3$.
Geological survey information was supplied by Cleveland
Potash Ltd, the operators of the Boulby mine.  
The average atomic number of Boulby rock was estimated to be
11.7$\pm$0.5 and the atomic mass to be 23.6$\pm$1.0.    The vertical
depth of the laboratory was measured as 1070~m and the density was
estimated to be 2.70$\pm$0.10~g/cm$^3$. 
It was found that the muon energy spectrum and mean muon energy at the
depth of the Boulby mine laboratory do not
depend strongly on either the rock composition or the exact depth.

The detector simulation was performed using custom written C++ code.
Five surfaces are simulated to represent the geometry of the detector
(Figure \ref{fig:z1veto}).  The bottom of the detector is represented by two
hemispheres, the sides by two cylinders and the top by a hollow disc.
The phototubes are represented by discs within the hollow disc surface.

\begin{figure}[htb]
\begin{center}
\epsfig{figure=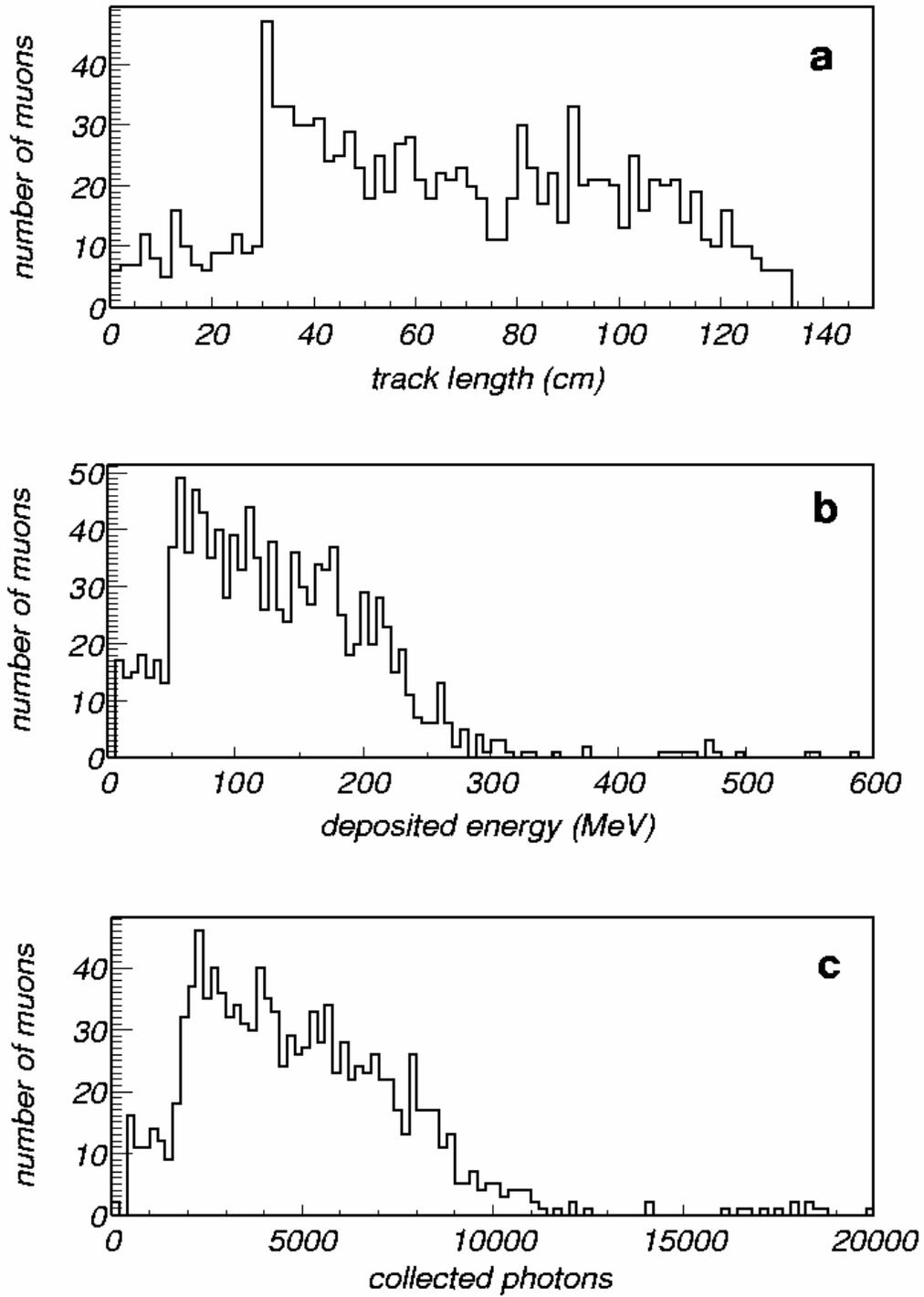,height=20cm}
\caption{Histograms of muon track length (a), energy loss (b) and collected
  photons (c) from simulation of muons passing through the detector}
\label{fig:hists}
\end{center}
\end{figure}

Muons are generated on the surface of a 2~m sided cube
centred on the origin of the hemisphere surfaces. 
Muon energy and direction were chosen according to the calculated
energy spectrum and angular distribution at a depth of 3000 metres of
water equivalent in Boulby rock, and the area of the detector
perpendicular to the muon flux was taken into account.
The assumption
that the muon direction is 
unaffected by interaction with the detector is used throughout the
simulation.  Once generated, 
each muon is propagated through the 2~m cube using a ray-tracing
algorithm until it passes through one of the five surfaces.  Once it
has passed through the surface, the muon is considered to be inside
the detector.  The muon is then propagated again until it passes through
another detector surface and is then considered to have left the detector.
Propagation of the muon continues until the ray-tracing algorithm
determines that it will not pass through any of the five surfaces again.
Due to the geometry of the veto, it is legitimate for the muon to pass
through the surfaces zero, two or four times.  This is a useful verification
employed in the code.  Further verification of the performance of the
simulation is provided by a GTk (\ul{G}IMP \ul{T}ool\ul{k}it) \cite{bib:gtk} graphical
user interface which uses the same ray-tracing algorithms used in the 
simulation to draw the veto surfaces, phototube surfaces and muon 
tracks to the screen.  The graphical user interface also serves to
confirm that the normals to the surfaces are correctly calculated
which is required for light transport (discussed below).
 The information acquired by propagating the
simulated muons is used to calculate the depth
of scintillator each muon passed through.  This is then used to
generate a track length histogram for a suitably large number of
simulated muons (Figure \ref{fig:hists}a).

Since the thickness of the scintillator is small, all processes of
muon interaction with matter other than ionisation could be neglected.
Muon energy deposition due to ionisation in the detector is sampled from a
Landau distribution.  The mean energy loss for ionisation and atom
excitation is calculated from the muon energy, path length
through the scintillator and the properties of the scintillator (Table
\ref{tab:scintillator}) using the Bethe-Bloch formula 
\cite{bib:bethebloch}.  A second histogram was generated showing
the distribution of energy deposition (Figure \ref{fig:hists}b).

The calculations up to this point were verified by repeating the
simulation using the FLUKA \cite{bib:fluka} code and
comparing the distributions of energy deposition. 

Due to the position of the phototubes on the detector and the overall
detector shape, light collection
for light emitted in the bottom of the detector is less efficient than
for light emitted in the sides.  This is clear from the fact that
there is a direct line of sight from the phototubes to the sides of the
detector but not to the bottom.  In order to account for this, it was
necessary for the simulation to also include light transport of the
scintillation photons within the detector.  Once the energy deposition of
each muon has been calculated, that value is converted into
a number of photons decided randomly based on a Gaussian
distribution.  The mean of the Gaussian distribution is calculated
from the muon energy deposition and mean scintillation
efficiency (Table \ref{tab:scintillator}).  The square root of the
mean is used as the standard deviation of the Gaussian distribution.
The initial position of each photon is determined by choosing a random
point along the muon track through the veto and the direction is
chosen randomly from a uniform solid angle distribution.

Photon transport is handled by repeatedly reflecting each photon from 
the five detector surfaces until its fate is decided.  After photon
initialisation and following each reflection, the same ray-tracing algorithms
used in muon propagation are used to determine the next surface which
the photon will hit.  The probability of bulk absorption within the
liquid along the path
to that surface is calculated based on the scintillator attenuation
length (Table \ref{tab:scintillator}) and a random number generated to
determine,  based on that probability, whether the photon should be
considered absorbed.  For the purposes of photon reflection, the
surfaces are modelled as slightly crumpled metal.  To simulate this,
the normal to the surface is rotated by a random
amount within a limited range.  Upon each intercept with a detector
surface, the photon is reflected specularly from the surface based on
the modified normal to the surface.  The probability of photon
absorption on the surface is calculated from the reflectance of
the surface and a random number is generated to determine
whether the photon should be considered absorbed or
reflected.  This process continues until the photon undergoes
absorption on a surface, bulk absorption in the liquid or intercepts the
photocathode of a photomultiplier.  If the photon intercepts a
photocathode before being absorbed, a random number is generated to
determine, based on the PMT efficiency (taken as 20\%), whether the photon
is detected.  If a photon intercepts a photocathode and is detected,
this is recorded.  Any photon hitting a photocathode and not being
detected is considered to have been absorbed.  In this way a histogram
of number of detected photons is built up (Figure \ref{fig:hists}c)
which, with suitable normalisation, may be compared to the measured spectrum. 

\section{Results}

Measurements were taken of the rate of muons depositing 30~MeV or more
in the detector over 53~days.  The data were taken over three runs,
the results of which are given in Table \ref{tab:results}.

\begin{figure}[htb]
\begin{center}
\epsfig{figure=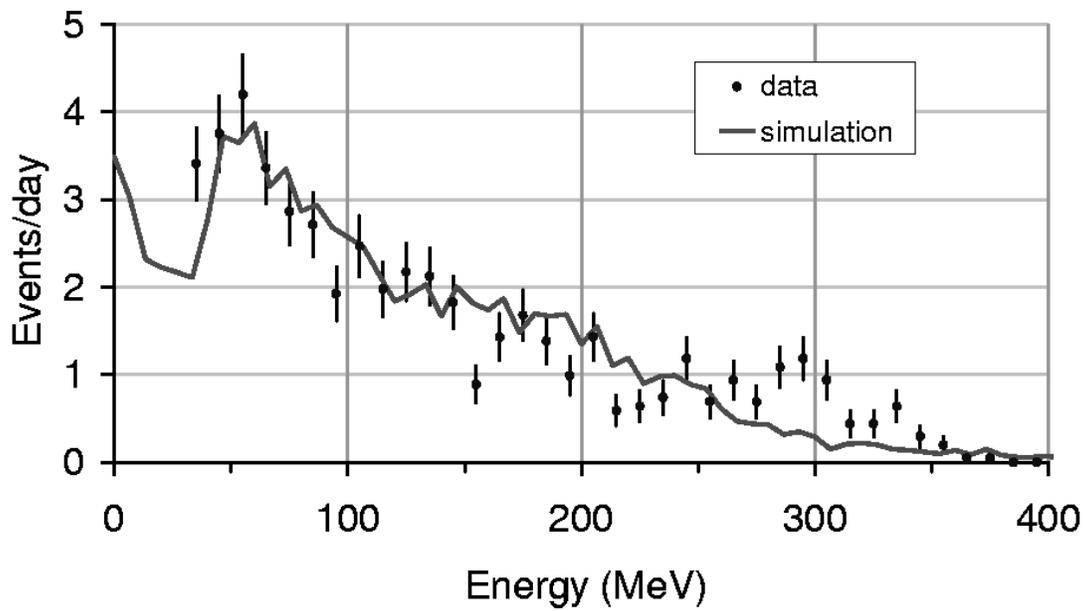,width=15.5cm}
\caption{Muon energy deposition spectrum using data collected in the
run starting 11-9-02 (Table \ref{tab:results}).}
\label{fig:spectra}
\end{center}
\end{figure}

Figure \ref{fig:spectra} shows the energy spectrum for the run dated
11th~September compared with the normalised
simulated spectrum of energy deposition.  The spectrum of energy
deposition was obtained from the spectrum of collected photons using
the calibration peak at around 60~MeV discussed above.

It was possible to use the simulation to calculate the proportion of
muons passing through a 2~m sided cube around the veto which can be
expected to deposit 30~MeV or more in the detector.
These figures were then used to calculate the flux of muons in the
laboratory.  Equation \ref{eq:area} gives the effective area $\langle$S$\rangle$ of
a 2~m sided cube normalised by relative muon intensity:
 
\begin{equation}
\label{eq:area}
\langle S\rangle=\frac{\int{I_\mu(\theta,\phi)S_\bot(\theta,\phi)d\Omega}}{\int{I_\mu(\theta,\phi)d\Omega}}
\end{equation}

where $I_\mu(\theta,\phi)$ is the flux of muons as a function of
zenith angle ($\theta$) and azimuthal angle ($\phi$) in
the laboratory, $S_\bot(\theta,\phi)$ is the area of the 2~m sided cube
perpendicular to the muon flux as a function of zenith and azimuthal angle.
At the Boulby laboratory depth, this calculation gives an effective area of
58029~cm$^2$.  The simulations showed that of those muons passing
through the 2~m sided cube, a fraction 0.261$\pm$0.003(stat) are expected to
pass through the detector and deposit 30~MeV or more in the detector.
This figure gives an
effective area for the detector of 15146~cm$^2$.
The calculated effective area was found to depend only slightly on the
estimate of the vertical depth of the laboratory. Reducing the
estimated depth by 200~m~w.e., the effective area was
calculated as 15139~cm$^2$.  (Note that the standard analysis
technique involving the evaluation of the detector acceptance as
a function of $\theta$ and $\phi$, described in references 6 and 8,
cannot be applied because the information about the muon directions is
not available.  The angular bins to be included in the calculation of
the acceptance are therefore not known.)
Using the total rate from Table \ref{tab:results}, the flux is calculated
to be $I_\mu=\rm(4.09\pm0.08(stat)\pm0.13(syst))\times10^{-8}\
 cm^{-2}s^{-1}$.
The systematic error is made up of contributions from cuts applied to
the measured spectra to remove $\gamma$ background, contributions from 
the simulations, uncertainty in the non-linearity correction applied
to the data, and uncertainty in the energy calibration of the
measured spectrum. 
Assuming a flat surface, the effective vertical depth of the laboratory in terms of
metres of water equivalent
was estimated by comparing the calculated muon intensities (See
Section \ref{sec:sim} for a description of the simulations) with our
measurements.
Using atomic number 11.7 and atomic mass 23.6 as
discussed in Section \ref{sec:sim} for Boulby rock, this was
calculated as 2805$\pm$15~m~w$.$e$.$
based on the measured muon flux.  Using
atomic number 11 and atomic mass 22 as for standard rock, the effective vertical depth would be
2845$\pm$15~m~w$.$e.  Based on the 
difference between the two vertical depth calculations the systematic
error on the vertical depth due to incomplete information on the rock
composition is estimated to be $\pm$40~m~w$.$e.  The final vertical
depth estimate is therefore 2805$\pm$45~m~w$.$e$.$.
Using the effective
vertical depth and the measured depth of the laboratory (1070~m), the mean rock
density is calculated as 2.62$\pm$0.01~g/cm$^3$ for Boulby rock and
2.66$\pm$0.01~g/cm$^3$ for standard rock.

For purposes of comparison with other similar measurements, the
vertical muon intensity was calculated:

\begin{equation}
\label{eq:vert}
I_{vert}=\frac{N_\mu}{\langle S\Omega\rangle\epsilon}=\rm 3.32\times10^{-8}\space cm^{-2}sr^{-1}s^{-1}
\end{equation}

Where $I_{vert}$ is the vertical muon intensity, $N_\mu$ is the muon rate in the detector
in s$^{-1}$, $\langle S\Omega\rangle$ is the
angular acceptance of the detector in cm$^2$sr and
$\epsilon$ is the fraction of muons depositing $>$30~MeV in the
detector calculated in the simulations.  The quantity $\langle
S\Omega\rangle$ was calculated from the simulations in a similar way
to the effective area $\langle S\rangle$ (See Equation \ref{eq:area}).

\section{Conclusions}

Muon rate data were collected using the \zeplin\ I veto detector in
the Boulby mine laboratory.
These data were used in conjunction with simulations of the detector
and of the muon spectrum to
calculate the muon flux in the laboratory.

The muon flux in the laboratory has been
measured as $(4.09\pm0.15)\times10^{-8}$~cm$^{-2}$s$^{-1}$.  This shows an effective
vertical depth of 2805$\pm$45 metres of water equivalent, and a mean rock
density of 2.62$\pm$0.03~g/cm$^3$.

These measurements will provide normalisation to neutron flux
simulations and help to design muon veto systems for future
dark matter experiments.

\section{Acknowledgements}

The authors would like to thank the members of the UK Dark Matter
Collaboration for their valuable assistance and advice.  
We are grateful to the Particle Physics and
Astronomy Research Council for financial support and to Cleveland Potash
Limited for their assistance.
M. Robinson would also like to thank Hilger Crystals for their support
of his PhD work.

\pagebreak

\begin{table}[htb]
\begin{center}
\caption{Specifications of liquid scintillator EJ-399-01}
\label{tab:scintillator}
\vspace{3mm}
\begin{tabular}{|l|l|}
\hline
Density & 0.89~g/cm$\rm^3$ \\
\hline
Wavelength of maximum emission & 425~nm \\
\hline
Light output & 57\% Anthracene \\
\hline
Attenuation length & $>$2~m \\
\hline
Flash point & 145$^\circ$C \\
\hline
Average atomic number & 4.75 \\
\hline
Average atomic mass & 8.33 \\
\hline
\end{tabular}
\end{center}
\end{table}

\begin{table}[htb]
\begin{center}
\caption{Muon rates in data collection runs} 
\vspace{3mm}
\begin{tabular}{|c|c|c|c|}
\hline
\label{tab:results}
run start & N$_{muons}$(E$>$30~MeV) & duration (days) &
rate (day$^{-1}$)\cr
\hline
26-7-02 & 1028 & 19.830 & 51.8$\pm$1.6(stat) \cr
28-8-02 & 712 & 12.886 & 55.3$\pm$2.1(stat) \cr
11-9-02 & 1097 & 20.249 & 54.2$\pm$1.6(stat) \cr
\hline
Total & 2837 & 52.965 & 53.6$\pm$1.0(stat) \cr
\hline
\end{tabular} 
\end{center} 
\end{table}

\pagebreak

\end{document}